\documentclass[twoside]{article}
\usepackage{fleqn,espcrc2}

\usepackage{graphicx}

\newcommand{\AmS}{{\protect\the\textfont2
  A\kern-.1667em\lower.5ex\hbox{M}\kern-.125emS}}

\title{Influence of excited electron lifetimes on the electronic structure of carbon nanotubes}

\author{Tobias Hertel\thanks{Corresponding author. Tel.: +49 30 8413-5505; fax: +49 30 8413-3383;
e-mail: hertel@fhi-berlin.mpg.de.}\address{Fritz--Haber--Institut der
Max--Planck--Gesellschaft, Faradayweg 4--6, 14195 Berlin, Germany} and Gunnar Moos }

\begin{document}

\begin{abstract}
We have studied the dynamics of electrons in single wall carbon nanotubes using
femtosecond time-resolved photoemission.  The lifetime of electrons excited to the $\pi
^{\ast}$ bands is found to decrease continuously from 130\,fs at 0.2\,eV down to less
than 20\,fs at energies above 1.5\,eV with respect to the Fermi level.  This should lead
to a significant lifetime--induced broadening of the characteristic van Hove
singularities in the nanotube DOS. \vspace{1pc}
\end{abstract}

\maketitle

\section{Introduction}

The unique electronic and mechanical properties of carbon nanotubes have stimulated
considerable interest in these materials with their possible applications as
nano--electronic devices or in new composite materials \cite{saito,ebbesen}. One of the
outstanding features of individual single wall carbon nanotubes (SWNTs) is their quasi
one--dimensional (1D) band--structure which can be derived by zone folding the
two--dimensional band--structure of graphene onto the 1D Brillouin zone of the tubes
\cite{saito}. The nanotube density of states (DOS) then exhibits a series of
characteristic van Hove singularities (VHS) that diverge like $E^{-1/2}$. Recent
experiments using scanning tunneling spectroscopy to study the electronic structure of
SWNTs were indeed able to resolve features in the tunneling spectra that are
characteristic for 1D van Hove singularities \cite{wildoer,odom,kim}. The width and
position of these features, however, is modified with respect to the results from
idealized tight binding calculations. It is generally accepted that interactions of the
nanotubes with their environment, {\it e.g.} tube--tube and tube--substrate interactions,
will lead to a deviation from 1D behaviour and will induce modifications in the SWNT band
structure. The scattering of electrons with phonons or other electrons will also modify
the 1D band structure and transport properties of SWNTs. Such scattering processes result
in a finite lifetime of excited electrons which will be investigated here. The observed
short electron lifetimes are expected to lead to a broadening of VHS. Other effects
contributing to a shift and broadening of features in the nanotube DOS are
curvature--induced $\sigma$--$\pi$ hybridization \cite{blase} and bond disorder
\cite{crespi}. Knowledge of the influence of different effects for changes of the
electronic structure, therefore, appears to be crucial for the understanding of the
spectroscopy of SWNTs.
\par
Here we present the first time--domain study of electronic excitations in SWNTs using
femtosecond time--resolved photoemission. We have studied the room temperature electron
dynamics at energies between -0.1\,eV and 2.3\,eV with respect to the Fermi level --- a
region which is of considerable interest in tunneling spectroscopy but of which only a
fraction is accessible by conventional transport studies. The measured electron lifetimes
should contribute significantly to the broadening of VHS in the nanotube DOS. We discuss
the implications of these results for the spectroscopy of SWNTs.

\section{Experimental}

Single--wall carbon nanotube samples used in this study are made from as--produced soot
and from commercial nanotube suspension (tubes@rice, Houston, Texas) with similar results
for both types of samples. The as--produced soot is pressed mildly between two glass
slides with a pressure of about 6\,kg\,cm$^{-2}$ to form a mat of entangled nanotube
ropes. The commercial nanotube suspension, containing SWNTs whose diameter distribution
is sharply peaked at 12\,$\rm \AA$, is used to fabricate bucky paper samples according to
the procedure described in reference \cite{rinzler}. The chiral wrapping angle which also
plays a crucial role for the electronic structure was recently found to vary widely
\cite{wildoer,odom} leading to a mixture of conducting and semiconducting tubes within
these samples. The samples are attached to a tantalum block which can be resistively
heated up to $1200^{\rm o}\,\rm C$. They are outgassed thoroughly by repeated heating and
annealing cycles under ultra high vacuum conditions. Photoelectron spectra are obtained
by means of the time of flight technique with an energy resolution of $\approx 10\,{\rm
meV}$. The visible pump pulses with a photon energy of typically 2.32\,eV are focused
onto the sample nearly collinearly together with the frequency doubled UV probe pulses
(pulse width 85\,fs). The beam waist at the sample position is 50\,$\mu$m in diameter.
More details about the experimental setup can be found in reference\,\cite{knoesel}. The
experiments were performed on a number of samples and on different spots on each sample
to ensure reproducibility of the data.

\begin{figure}[htb]
\centering
\includegraphics[width=\columnwidth]{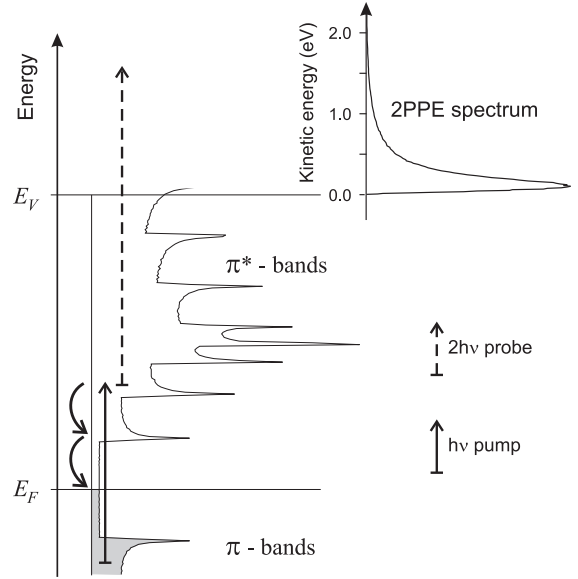} \caption{Schematic illustration of the pump--probe scheme employed in our
experiments. Initially a visible femtosecond laser pulse excites electrons from the $\pi$
to the $\pi ^{\ast}$ bands. The resulting electron dynamics is monitored with a second UV
probe pulse which photoemits electrons into the vacuum after a well defined time delay. A
typical photoelectron spectrum is shown in the upper right of the figure. \label{fig1}}
\end{figure}

\section{Results and discussion}

The visible--pump UV--probe scheme employed in our experiments is illustrated
schematically in Fig.\,\ref{fig1} (for a review of the time--resolved photoemission
technique see reference \cite{ogawa}). The pump pulse initially excites electrons from
the occupied $\pi$ bands to some intermediate state of the unoccupied $\pi^{\ast}$ bands
above the Fermi level. The electrons then loose energy by various scattering processes
and relax towards the Fermi level. After a well defined time--delay a second UV probe
pulse photoemits electrons into the vacuum where they are detected energy selectively. If
the photoemission signal at a certain energy is recorded as a function of the time--delay
we obtain cross--correlation traces which directly reflect the electron dynamics. The
intermediate state energy of the probed electrons $(E-E_{F})$ is calculated from the
electron kinetic energy $E_{kin}$ via the relation $(E-E_{F})= E_{kin} + e\Phi -
h\nu_{probe}$ , where $e\Phi=4.52\pm 0.05\,{\rm eV}$ is the sample work--function.

\par
The 2PPE spectrum also shown in Fig.\,\ref{fig1} does not exhibit any signs of the VHS
characteristic for the band structure of SWNTs. We attribute this to a combination of
effects: a) The width of the diameter distribution of nanotubes in the sample gives rise
to VHS at different energetic positions in the corresponding density of states. b)
Tube--tube interactions may lead to a broadening of features in the DOS. Calculations by
different authors indicate that tube-tube interactions may cause a band splitting of
0.1\,eV to 0.5\,eV for nanotube arrays \cite{delaney,charlier,kwon}. c) Most importantly
it needs to be considered that
--- unlike optical absorption \cite{petit} or EELS \cite{pichler} measurements --- photoemission is very sensitive to
the position of the band structure with respect to the vacuum level. Shifts in the
alignment of the band structure can be induced by charge transfer between different tube
species ('self--doping'). In particular the latter effect should contribute to the
smearing of VHS beyond recognition in our 2PPE spectra and indicates the presence of
tube--tube interactions. Here, we take advantage of one of the particular strengths of
time--resolved techniques: they allow to study the dynamics in systems with
inhomogeneously broadened or heavily congested spectra.

\begin{figure}[thb]
\includegraphics[width=\columnwidth]{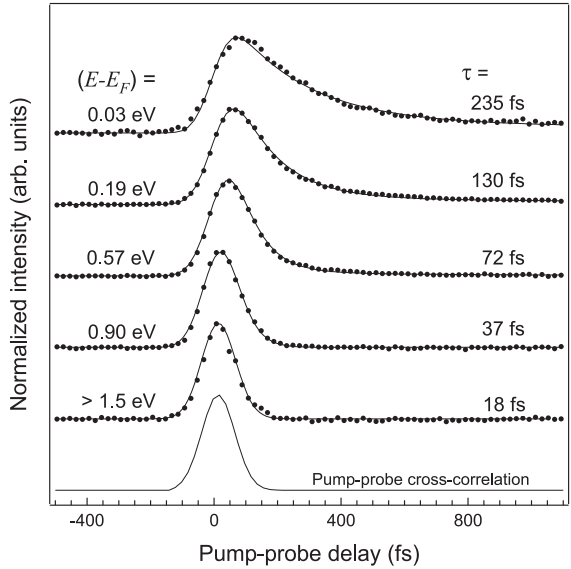}
\caption{Cross--correlation traces for various intermediate state energies. The electron
dynamics are obtained from the exponential decay of the correlated signal. \label{fig2}}
\end{figure}

\par
\begin{figure}[thb]
\includegraphics[width=\columnwidth]{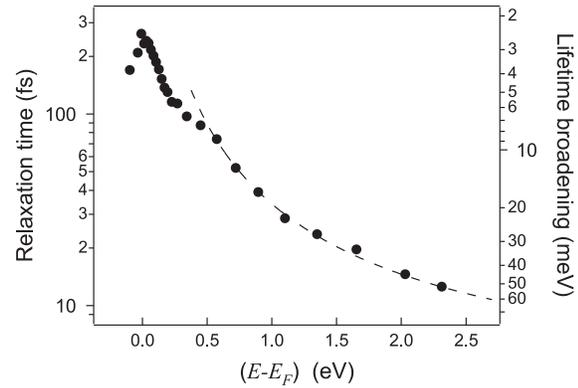} \caption{Relaxation time obtained
from the exponential fit to cross--correlations for various intermediate state energies.
The corresponding lifetime broadening is given on the right axis. This fast decay may be
attributed to electron--electron interactions and decreases continuously to less than
20\,fs at energies above 1.5\,eV. The dashed line is an empirical fit to the higher
energy data and is used for the computation of the lifetime broadened spectra in
Fig.\,\ref{fig4}.\label{fig3}}
\end{figure}

Typical cross-correlation traces recorded with a narrow energy pass of about 50\,meV are
shown in Fig.\,\ref{fig2} for various intermediate state energies. The cross-correlations
can be fit with a single exponential decay except for intermediate state energies close
to the Fermi level where we also find a small contribution from a slower component which
decays on the picosecond time--scale. The slow decay is attributed to the cooling of the
laser heated electron gas after the initial fast electron dynamics have lead to a
thermalization of the excited electron distribution and will be discussed elsewhere
\cite{hertelprep}. The decay--time of the fast component can be seen to increase
continuously as the intermediate state energy approaches the Fermi--level (see Fig.
\ref{fig3}).

\par The detailed kinetics of the electron relaxation involves the decay
from a particular exited state as well as the filling of the same state by electrons
which themselves decay from higher lying states. The refilling becomes more important at
energies close to the Fermi level and leads to an overestimation of the electron
lifetimes by the corresponding cross--correlation decay. However, an estimate of the
influence of such secondary electron cascades on the 2PPE signal reveals that the actual
lifetime at 0.2\,eV above the Fermi level is at most 20\% shorter than the measured decay
time \cite{comment}. We, therefore, assume that the influence of secondary electron
cascades on the measured decay times at energies above about 0.2\,eV represents a minor
correction to the actual lifetimes. Irrespective of the magnitude of the effect, however,
the measured decay times used to estimate the lifetime--induced broadening of VHS should
give a lower bound for the actual broadening.

\par
A qualitatively similar behaviour is also observed in simple metals where the electron
lifetime ideally increases with $(E-E_{\rm F})^{-2}$ due to phase space limitations for
scattering events near the Fermi level (Fermi--liquid behavior) \cite{ogawa,hertel}. In
ideal one--dimensional systems the energy dependence of the electron lifetimes is
expected to be proportional to $(E-E_{\rm F})^{n}$ with $n=1$ \cite{luttinger}. In our
study the energy dependence of the cross--correlation decay $\tau$ (at energies above
0.5\,eV) can best be approximated by the empirical function $\tau=[30((E-E_{\rm F})/{\rm
eV})^{-1.5}+4]$\,fs. We note, however, that the energy dependence of electron lifetimes
should only be used to draw conclusions with respect to the character of the system ---
{\it e.g.} Luttinger--liquid {\it vs} Fermi--liquid --- in combination with {\it ab
initio} studies that account for all band structure effects including the interaction
with the nanotube environment. Recent calculations of the electron dynamics in aluminum,
for example, revealed that band structure effects can lead to a substantial deviation of
electron lifetimes from the Fermi--liquid predictions \cite{campillo,schoene} in spite of
the fact that aluminum is generally considered to be a nearly ideal free electron gas. In
addition we note that the electron lifetimes in 2--D graphite are also expected to scale
with an exponent $n$ close to unity \cite{gonzalez}.

\begin{figure}[htb]
\includegraphics[width=\columnwidth]{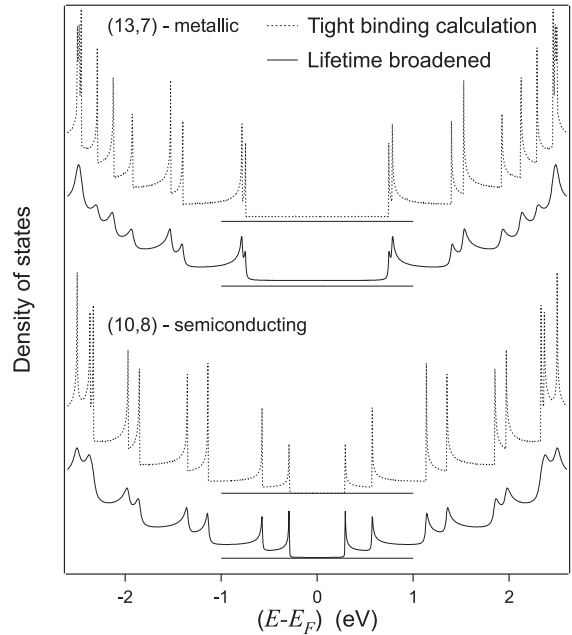} \caption{Influence of the electron
lifetime on the calculated nanotube DOS for the metallic $(13,7)$ and the semiconducting
$(10,8)$ species. The upper curve of each pair of traces gives the DOS obtained from
tight binding calculations while the lower traces are obtained by a convolution of that
DOS with an energy dependent Lorentzian. The van Hove singularities are seen to be
increasingly broadened at higher energies due to the strong decrease of the electron
lifetime with energy.\label{fig4}}
\end{figure}

\par
The short decay--times observed here suggest that the fast energy relaxation dynamics in
nanotubes is likely due to electron--electron ($e$--$e$) scattering in analogy to recent
experimental studies and calculations on $e$--$e$ interactions in graphite
\cite{xu,gonzalez}. Electron--phonon ($e$--$ph$) scattering is expected to be much slower
and would provide less efficient energy transfer for the scattering electrons
\cite{jishi}.

\par
These finite electron lifetimes lead to a modification of electronic spectra with respect
to the spectra expected for an isolated, non--interacting system. We illustrate the
effect this has on the electronic structure of carbon nanotubes by convoluting the tight
binding density of states $N(E)$ with a Lorentzian of energy dependent width $\Gamma(E)$
(see Fig.\,\ref{fig4}). The energy dependence of the line--width is obtained from the
empirical fit to the lifetimes $\tau$ in Fig.\,\ref{fig3} according to $\Gamma=\hbar
/\tau$, where we assume identical lifetimes for electron and hole excitations. Note that
the observed decrease of the lifetime to $\approx$\,20\,fs at an energy of 1.5\,eV leads
to a broadening of features in the SWNT DOS of 30\,meV. At still higher energies this
presents only a lower limit to the expected lifetime broadening as given by the time
resolution of these experiments. In Fig.\,\ref{fig4} we show the results of this
convolution for the metallic (13,7) tube (recently identified by tunneling spectroscopy
\cite{kim}) and the semiconducting (10,8) tube. The density of states for the $\pi$ and
$\pi^{\ast}$ derived bands was obtained in the usual way by zone folding the 2D graphite
band structure into the 1D Brillouin zone of the nanotubes with a nearest neighbor
overlap integral $\gamma_{0}$ of 2.5\,eV. It is evident that the influence of the
electron lifetime on the shape of the VHS in the DOS becomes more pronounced at higher
energies and may even lead to the coalescence of VHS. The magnitude of this effect is
about a factor of 2--10 smaller than the band splitting calculated for nanotube arrays
\cite{delaney,charlier,kwon} but should still give a significant contribution to the
electronic spectra in particular at higher electron energies.

\par
Another interesting aspect of this study is that we do not observe any slow decay that
can be assigned to carrier recombination processes across the band gap of semiconducting
tubes, despite the abundance of such tubes in these samples. Interband recombination is
generally expected to be quite slow like in ${\rm C}_{60}$ films
(30\,ps\,--\,40\,ps)\cite{fleischer} or in silicon where recombination occurs on the ns
time--scale \cite{goldman}. We take the absence of such a slow channel as further
evidence that tube-tube interactions are sufficiently strong to induce charge transfer
between semiconducting and metallic tubes on a time--scale comparable to or faster than
the observed decay. Actually the lifetimes found here are qualitatively similar to
results obtained for highly oriented pyrolytic graphite where the decay is found to be
about a factor of 1.5-2 slower than in the nanotube samples \cite{hertelprep}. Note that
tube--tube interactions which lead to only 50\,meV band--shift or splitting would already
allow charge transfer between tubes within $\hbar / 50\,{\rm meV}\approx 13\,{\rm fs}$.
Charge transfer between different tube types may also be enhanced by scattering from
static or dynamic lattice distortions. Also note that the aforementioned slow channel
from the cooling dynamics cannot be assigned to the semiconducting species since it only
contributes to the 2PPE signal at energies below about 0.3\,eV where no significant
photoemission is expected from semiconducting tubes in these samples (the average
band--gap is about 0.55\,eV).
\par
Electron--electron scattering may provide an important mechanism for electron phase
relaxation
--- in particular in low temperature transport studies. If compared with calculated
$e$--$ph$ scattering times of about 1.4\,ps \cite{jishi} the short electron lifetimes
observed in this study suggest that $e$--$e$ scattering may dominate phase relaxation
even at room temperature if electron energies of a few hundred meV above the Fermi level
are considered. The corresponding phase relaxation lengths can be estimated from the data
of Fig.\,\ref{fig3}. This yields the energy dependence of the $e$--$e$ contribution to
the phase relaxation length $L_{\varphi}=[ 3^{-1/2}\,v_{F}\,\tau_{e-e}]$ which increases
from less than 10\,nm at energies above 1.5\,eV up to about 60\,nm at an energy of
0.2\,eV ($v_{\rm F}= 8\times10^{5}\,{\rm m\,s^{-1}}$, \cite{saito}). We note that at
energies below about 0.2\,eV $e$--$e$ scattering times cannot be determined unambiguously
from these experiments because a combination of $e$--$ph$ scattering and electron cascade
effects makes the interpretation of the data more difficult.
\par
In summary, we have characterized the energy dependence of the electron energy relaxation
time and illustrated the consequences for the electronic structure of SWNTs. The
continuous decrease of the electron lifetime to less than 20\,fs at energies above
1.5\,eV with respect to the Fermi level should lead to a significant broadening of the
VHS in the nanotube DOS. These findings should help to analyze spectra from carbon
nanotubes with respect to different contributions from various line--broadening
mechanisms. In order to predict the influence of the finite electron lifetime on spectral
features at higher energies it is desirable to obtain data with higher time--resolution
than in the present study. Ideally, phase--controlled measurements may provide very
detailed information on electron decoherence \cite{petek}. Our studies furthermore
indicate that tube--tube interactions in nanotube ropes are strong enough to lead to
electron transfer between neighboring metallic and semiconducting tubes on the
sub--picosecond time--scale.

\section*{Acknowledgements }
We thank R.E.\ Smalley and H.\ Dai for providing the unprocessed SWNTs. We acknowledge
stimulating discussions with M. Wolf. It is our pleasure to thank G.\ Ertl for his
continuing and generous support.


\begin{thebibliography}{9}

\bibitem{saito} R.\ Saito, G.\ Dresselhaus and M.\ Dresselhaus,
{\it Physical Properties of Carbon Nanotubes}, London, Imperial College Press, 1998.
\bibitem{ebbesen} T.\ W.\ Ebbesen,
{\it Carbon Nanotubes, Preparation and Properties}, Boca Raton, CRC Press, 1997.
\bibitem{wildoer} J.W.G.\ Wild\"{o}er, L.\ C.\ Venema, A.\ G.\ Rinzler, R.\ E.\ Smalley and C.\ Dekker, Nature\ {\bf 391}, 59 (1998).
\bibitem{odom} T.W.\ Odom, J.\-L.\ Huang, P.\ Kim and C.\ Lieber, Nature\ {\bf 391}, 62 (1998).
\bibitem{kim} P.\ Kim, T.\ W.\ Odom, J.\-L.\ Huang and C.\ M.\ Lieber, Phys.\ Rev.\ Lett.\
{\bf 82}, 1225 (1999).
\bibitem{blase} X.\ Blase, L.\ X.\ Benedict, E.\ L.\ Shirley and S.\ G.\ Louie, Phys.\ Rev.\ Lett.\
{\bf 72}, 1878 (1994).
\bibitem{crespi} V.H.\ Crespi and M.\ L.\ Cohen, Phys.\ Rev.\ Lett.\
{\bf 79}, 2093 (1997).
\bibitem{rinzler} A.\ G.\ Rinzler, J.\ Liu, H.\ Dai, P.\ Nikolaev, C.\ B.\ Huffman, F.\ J.\ Rodriguez-Macias, P.\ J.\ Boul, A.\ H.\  Lu, D.\ Heymann, D.\ T.\ Colbert, R.\ S.\ Lee, J.\ E.\ Fischer, A.\ M.\ Rao, P.\ C.\ Eklund and R.\ E.\ Smalley, Appl.\ Phys.\ A\ {\bf
67}, 29 (1998).
\bibitem{knoesel} E.\ Knoesel, A.\ Hotzel, and M.\ Wolf, Phys.\ Rev.\ B\ {\bf
57}, 12812 (1998).
\bibitem{ogawa} S.\ Ogawa and H.\ Petek, Prog.\ Surf.\ Sci.\ {\bf
56}, 239 (1998).
\bibitem{delaney} P.\ Delaney, H.\ J.\ Choi, J.\ Ihm, S.\ G.\ Louie and M.\ L.\ Cohen, Nature\
{\bf 391}, 466 (1998).
\bibitem{charlier} J.-C.\ Charlier, X.\ Gonze and J.\-P.\ Michenaud, Europhys.\ Lett.\ {\bf 29}, 43
(1995).
\bibitem{kwon} Y.-K.\ Kwon, S.\ Saito and D.\ Tom\'{a}nek, Phys.\ Rev.\ B {\bf 58}, R13314 (1998).
\bibitem{petit} P.\ Petit, C.\ Mathis, C.\ Journet and P.\ Bernier, Chem.\ Phys.\ Lett.\
{\bf 305}, 370 (1999).
\bibitem{pichler} T.\ Pichler, M.\ Knupfer, M.\ S.\ Golden, J.\ Fink, A.\ Rinzler, R.\
E.\ Smalley,  Phys.\ Rev.\ Lett.\ {\bf 80}, 4729 (1998).
\bibitem{hertelprep} T.\ Hertel {\it et al.}, in preparation.
\bibitem{comment} This estimate was obtained from a simulation of
the electron relaxation dynamics in a free electron metal including secondary electron
cascades. This complex process can be described analytically (see V.E.\ Gusev, and O.B.\
Wright, Phys.\ Rev.\ B\ {\bf 57}, 2878 (1998)) and was modeled for gold using parameters
appropriate to our experimental setup.
\bibitem{hertel} T.\ Hertel, E.\ Knoesel, M.\ Wolf and G.\ Ertl, Phys.\ Rev.\ Lett.\ {\bf
76}, 535 (1996).
\bibitem{luttinger} J.\ M.\ Luttinger, Phys.\ Rev.\ {\bf 121}, 942 (1961).
\bibitem{campillo} I.\ Campillo, J.\ M.\ Pitarke, A.\ Rubio, E.\ Zarate and P.\ M.\
Echenique, Phys.\ Rev.\ Lett.\ {\bf 83}, 2230 (1999).
\bibitem{schoene} W.\-D.\ Sch\"{o}ne, R.\ Keyling, M.\ Bandi\'{c} and W.\ Ekardt, Phys.\
Rev.\ B\ {\bf 60}, 8616 (1999).
\bibitem{gonzalez} J.\ Gonz\'{a}lez, F.\ Guinea and M.\ A.\ H.\ Vozmediano, Phys.\ Rev.\ Lett.\ {\bf 77}, 3589 (1996).
\bibitem{xu} S.\ Xu, J.\ Cao, C.\ C.\ Miller, D.\ A.\ Mantell, R.\ J.\ D.\ Miller and Y.\ Gao, Phys.\ Rev.\ Lett.\ {\bf 76}, 483 (1996).
\bibitem{jishi} R.A.\ Jishi, M.\ S.\ Dresselhaus and G.\ Dresselhaus, Phys.\ Rev.\ B\ {\bf
48}, 11385 (1993).
\bibitem{fleischer} S.B.\ Fleischer, E.\ P.\ Ippen, G.\ Dresselhaus, M.\ S.\ Dresselhaus, A.\ M.\ Rao, P.\ Zhou and P.\ C.\ Eklund, Appl.\ Phys.\ Lett.\ {\bf 62}, 3241 (1993).
\bibitem{goldman} J.R.\ Goldman and J.A.\ Prybyla, Phys.\ Rev.\ Lett.\ {\bf 72}, 1364 (1994).
\bibitem{petek} H.\ Petek, H.\ Nagano and S.\ Ogawa, Phys.\ Rev.\ Lett.\ {\bf 83}, 832
(1999).
\end{thebibliography}
\end{document}